\documentclass[twocolumn,showpacs,preprintnumbers,amsmath,amssymb]{revtex4}

\usepackage{amsmath,amssymb,graphicx}
\usepackage{graphicx}
\usepackage{dcolumn}
\usepackage{bm}
\usepackage{epsfig}
\usepackage{here}
\usepackage{float}
\usepackage{amsmath}
\usepackage[usenames]{color}

\newcommand{\bea}{\begin{eqnarray}}
\newcommand{\eea}{\end{eqnarray}}

\begin{document}
\draft

\title{Roles of dark energy perturbations in the dynamical dark energy models: \\
       Can we ignore them?}
\author{Chan-Gyung Park${}^{1}$, Jai-chan Hwang${}^{1}$, Jae-heon Lee${}^{1}$
        and Hyerim Noh${}^{2}$}
\address{${}^{1}$Department of Astronomy and Atmospheric Sciences,
                 Kyungpook National University, Taegu, Korea \\
         ${}^{2}$Korea Astronomy and Space Science Institute,
                 Daejon, Korea}


\begin{abstract}

We show the importance of properly including the perturbations of
the dark energy component in the dynamical dark energy models based
on a scalar field and modified gravity theories in order to meet
with present and future observational precisions. Based on a simple
scaling scalar field dark energy model, we show that observationally
distinguishable substantial differences appear by ignoring the dark
energy perturbation. By ignoring it the perturbed system of
equations becomes {\it inconsistent} and deviations in
(gauge-invariant) power spectra {\it depend on} the gauge choice.

\end{abstract}

\noindent 

\maketitle

%
%
The high-$z$ type Ia supernovae (SNIa) luminosity-distance relation
suggests that the expansion rate of our universe is currently under
acceleration \cite{SN}. The cosmological constant is readily
(re)introduced to explain the observation theoretically. Theoretical
studies of the large-scale structure formation process imprinted in
the matter power spectrum \cite{BAO,SDSS-DR7-LRG} and the cosmic
microwave background radiation (CMB) power spectrum \cite{WMAP-5yr}
also favor presence of substantial amount of agent with repulsive
nature like the cosmological constant. With the advent of the recent
acceleration, the long lasting age-problem of the world model, which
has persisted ever since the first observation of the expansion of
the universe, has now evaporated from the cosmological scene.

The nature of the agent causing the acceleration, however, is still
unknown and it is one of the fundamental mysteries in the present
day theoretical cosmology. Although the cosmological constant is a
historically well known possibility, it also has two well
appreciated problems: the cosmological constant (why so small)
problem and the coincidence (why now or fine tuning) problem. Large
amount of literature has been devoted to address these problems,
especially the latter one, by introducing dynamical agents, often
termed the dark energy. As far as we can tell the fine tuning
problem has not been properly addressed even using the dynamical
dark energy. Introduction of the dynamic possibility of the dark
energy, however, has opened a whole new arena for cosmological
research based on variety of possibilities using field, fluid,
modified gravity, other dimensions, etc.

In the case of the cosmological constant as the dark energy, due to
its constant nature (both in time and space) its contribution
directly appears {\it only} in the background world model. However,
when we consider the dynamical dark energy we should pay attention
to its dynamical roles not only in the background world model but
also in the structure formation process. Here, we address the
importance of properly including the role of dark energy
perturbation (DEP) imprinted in the large-scale matter and the CMB
anisotropies power spectra, and the perturbation growth process
especially in the context of present and future observations with
due precision.

Recent dramatic progresses made in the observational cosmology open
possibility to constrain the character of the dark energy, and call
for equally precise theoretical tools in the cosmic structure
formation process. The expansion history based on the SNIa, the
matter power spectrum, the CMB anisotropy power spectra, and the
perturbation growth factor provide four domains where theories meet
with observations. The relevant present and future observation
programs in the CMB, SNIa, and the large-scale clustering
include the WMAP (Wilkinson Microwave Anisotropy Probe) and the
Planck missions, 2dFGRS (Two-degree-Field Galaxy Redshift Survey),
SDSS (Sloan Digital Sky Survey), to mention a few. The large-scale
clustering can be probed by diverse observations: weak lensing,
Lyman-$\alpha$ and hydrogen 21cm tomography, X-ray galaxy clustering
mass function, galaxy redshift-space distortion, integrated
Sachs-Wolfe effect, etc. In the following we will compare our
results with SDSS DR7 (seventh Data Release) for the matter power
spectrum \cite{SDSS-DR7-LRG}, WMAP 5-year data for the CMB spectrum
\cite{WMAP-5yr}, and the future X-ray and weak lensing observations
of clusters using X-ray surveys for the perturbation growth factor
\cite{X-ray}.

Our study is motivated by often used practices in the literature
which ignore the DEP even in the case of dark energy models using
the scalar field or modified gravity theories, see
\cite{no-DE-pert}. That is, in the presence of a dynamical dark
energy it is {\it not} guaranteed to use the following
conventionally known equation \cite{Lifshitz-1946} \bea
   \ddot \delta_b
       + 2 H \dot \delta_b
       - 4 \pi G \delta \rho_b = 0,
   \label{ddot-delta-eq}
\eea which is true {\it only} for the cosmological constant as the
dark energy; $\delta_b \equiv \delta \rho_b/\rho_b$ is the relative
density fluctuation of baryon component, $H \equiv \dot a/a$, $a$ is
the cosmic scale factor, and an overdot denotes a time derivative.
In the presence of dynamical dark energy we have contributions from
the DEP in the right-hand side which are accompanied by a
second-order differential equation describing the equation of motion
of the perturbed dark energy. Even in modified gravity context, in
the literature, we often notice a similar equation replacing $G$ by
some effective $G_{\rm eff}$. Without proper (perhaps numerical)
verification such a simplification is hardly allowed mathematically
because it corresponds to replacing a second-order differential
equation by an algebraic coefficient (which is zero in the above
case); as we have $G_{\rm eff} = G$ in Einstein's gravity limit, if
such an approximation of ignoring the DEP is not allowed in
Einstein's gravity the same is true even in modified gravity
context.

Indeed it is always prudent and correct to include the DEP in
principle, but more relevant issue would be whether we could ignore
such accompanied fluctuations in practice. In this {\it Letter}, by
using a simple dynamical dark energy model based on a scalar field
we will show that the answer is negative even in Einstein's gravity;
for related works, see \cite{DE-pert}.

\begin{figure}
\begin{center}
\includegraphics[width=8.6cm]{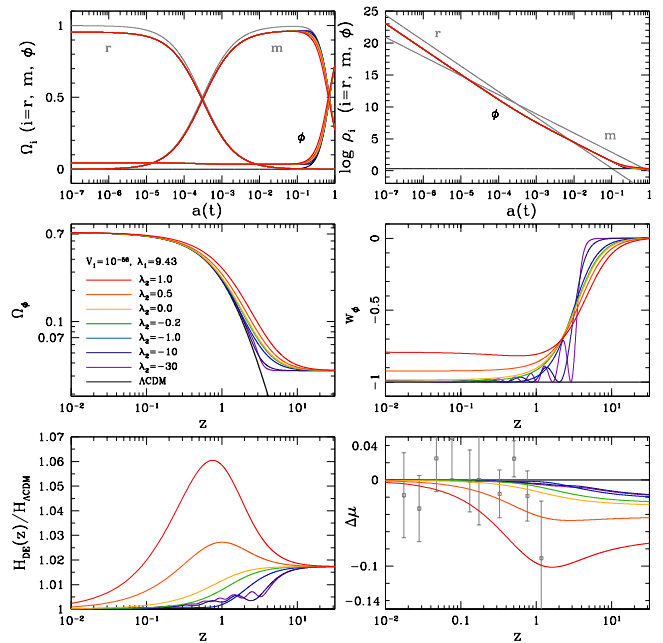}
\caption{
         Top panels: Evolution of $\Omega_i$ and $\rho_i$ as a function
         of scale factor $a(t)$ in the $\phi\textrm{CDM}$ universes with scalar field
         potential parameters set by Basset et al.\ \cite{Bassett-etal-2008}
         (colored curves),
         where  $i=\textrm{r},\textrm{m},\phi$ indicates radiation,
         matter (baryon + CDM),
         and scalar field, respectively. Black curves represent those of
         $\Lambda\textrm{CDM}$ model.
         Middle and bottom panels:  Evolution of $\Omega_\phi$, $w_\phi$,
         $H_{\textrm{DE}}(z)/H_{\Lambda\textrm{CDM}}$, and
         $\Delta\mu (z) = \mu_{\textrm{DE}}(z)-\mu_{\Lambda\textrm{CDM}}(z)$
         for the same set of $\phi\textrm{CDM}$ models.
         In the $\Delta\mu$-plot, the grey open squares with error bars
         represent the deviation of SNIa data points from the
         $\Lambda\textrm{CDM}$ model considered here.
         The binned SNIa data are based on the Union sample
         \cite{Kowalski-etal-2009}.
         }
\label{fig:BG-evolution}
\end{center}
\end{figure}

%
%
As a simple dynamical model of dark energy we consider a minimally
coupled scalar field with a double exponential potential (we set $c
\equiv 1 \equiv \hbar$)
   $ V(\phi) = V_1 e^{- \lambda_1 \phi}
       + V_2 e^{- \lambda_2 \phi} $,
where $\phi$ is the scalar field. The background evolution was
investigated previously by Bassett et al.\ in
\cite{Bassett-etal-2008}, and we consider the background parameters
of the scalar field suggested in that work: we call it a $\phi$CDM
(cold dark matter) model. As a fiducial model we take a flat
$\Lambda$CDM universe with parameters $\Omega_m = 0.274$
($\Omega_c=0.2284$ and $\Omega_b=0.0456$), $\Omega_\Lambda=0.726$,
$h=0.705$, $n_s=0.960$, $\sigma_8=0.812$, $T_0=2.725~\textrm{K}$,
$Y_\textrm{He}=0.24$, $N_\nu=3.04$ based on the WMAP 5-year
observations \cite{WMAP-5yr}, but without reionization. Evolution of
the background world models is presented in Fig.\
\ref{fig:BG-evolution}. For all $\phi\textrm{CDM}$ models we take
$V_1=10^{-56}$ and $\lambda_1=9.43$, and from red to violet curves,
$\lambda_2=1.0$, $0.5$, $0.0$, $-0.2$, $-1.0$, $-10$, and $-30$. For
each model, $V_2$ parameter has been determined to have the present
dark energy density parameter equal to $\Omega_\phi=0.726$. The
initial dark energy density parameter $\Omega_{\phi i}$ is
determined by the parameter $\lambda_1$; i.e., $\Omega_{\phi
i}=3(1+w)/\lambda_1^2=0.045$ during the radiation domination with
$w=\frac{1}{3}$, see Eq.\ (4) in \cite{HN-scaling-2001}.

Our dark energy model allows exact scaling during the radiation and
matter dominated eras (provided by $\lambda_1$-term) and behaves as
the dark energy in the present epoch (provided by $\lambda_2$-term).
Following \cite{Bassett-etal-2008} we consider the initial
contribution from the dark energy to be close to a maximum amount
allowed by the big bang nucleosynthesis (BBN) calculation
$\Omega_{\phi i} < 0.045$ \cite{Bean-etal-2001}. The parameters used
in our dark energy model are consistent with currently known
cosmological constraints from the BBN and the high-$z$ SNIa
observations, see Fig.\ \ref{fig:BG-evolution}.

%
%

%
%
\begin{figure}
\begin{center}
\includegraphics[width=8.6cm]{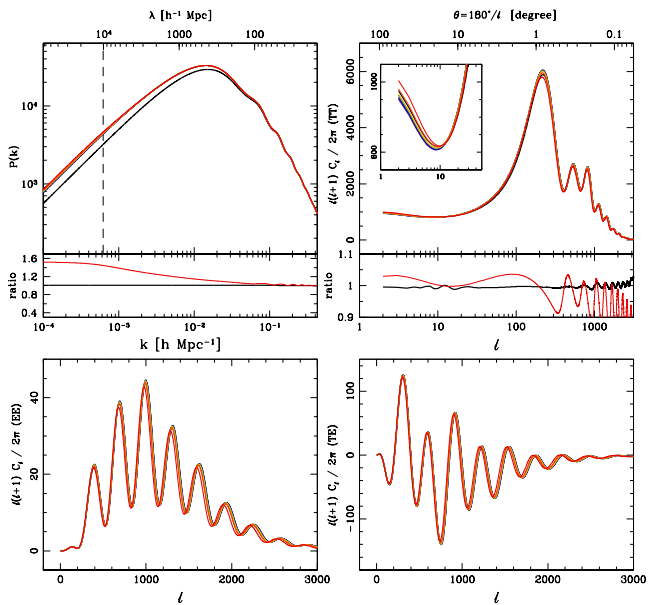}
\caption{
         The matter power spectrum (top-left), and CMB TT (top-right), EE
         (bottom-left), TE (bottom-right) power spectra of $\phi$CDM universe
         with scalar field potential parameters used in Fig.\
         \ref{fig:BG-evolution}, with the same colored code.
         Predictions of $\Lambda\textrm{CDM}$ model are shown as black curves.
         The vertical line in the top-left panel indicates the present horizon
         size ($10081~h^{-1}~\textrm{Mpc}$) of the
         $\Lambda\textrm{CDM}$ universe.
         The small box in the top-right panel magnifies the CMB TT powers
         at low $\ell$'s.
         All calculations are made in three different gauge conditions
         (CCG, UEG, and UCG), where evolution of perturbation of the dark
         energy scalar field has been properly considered.
         The results in the three gauges coincide exactly.
         The matter and CMB power spectra of the $\Lambda\textrm{CDM}$ model
         have been normalized with $\sigma_8$ and COBE spectrum, respectively.
         For comparison, all the $\phi\textrm{CDM}$ power spectra have been
         normalized with the $\Lambda\textrm{CDM}$ ones at small scales,
         $\ell=700$ for CMB and $k=0.3~h$ Mpc$^{-1}$ for matter ones.
         For a $\phi\textrm{CDM}$ with $\lambda_2=1.0$ that is most
         deviated from the $\Lambda\textrm{CDM}$ prediction, the ratios
         of its powers to our $\Lambda\textrm{CDM}$
         predictions are also shown in the bottom region of top
         panels;
         as an indication of numerical accuracy of our code
         ``the CMBFAST-derived power spectra \cite{Seljak-1996} divided
         by our result for $\Lambda\textrm{CDM}$ model'' is represented
         as black curve.
         }
\label{fig:pert-evolution-on}
\end{center}
\end{figure}

In order to calculate the matter and CMB power spectra, and
evolution of the baryon density perturbation we solve a system
composed of matter (dust and CDM), radiation (handled using the
Boltzmann equation or tight coupling approximation), together with
the cosmological constant or the scalar field as the dark energy.
Our set of equations and the numerical methods are presented in
\cite{HN-CMB-2002}. As the initial conditions for perturbation
variables we use the scaling solutions derived in
\cite{HN-scaling-2001}. We solved the system in three different
gauge conditions: the CDM-comoving gauge (CCG), the
uniform-expansion gauge (UEG), and the uniform-curvature gauge
(UCG); the CCG, the UEG, and the UCG, respectively, set the velocity
of the CDM, the perturbed expansion of normal frame vector (or the
perturbed trace of extrinsic curvature), and the perturbed part of
intrinsic scalar curvature equal to zero as the temporal gauge
condition; all perturbation variables we use are spatially
gauge-invariant \cite{Bardeen-1988}. The CCG is the same as the
synchronous gauge without the gauge mode. Each of these gauge
conditions fixes the gauge degrees of freedom completely, thus any
variable in these gauge conditions is equivalent to a unique
gauge-invariant combination of variables. Value of any
gauge-invariant variables evaluated in the three gauges should
coincide exactly. We used this to check the consistency of the
calculation and the numerical accuracy.

In Fig.\ \ref{fig:pert-evolution-on} we present the matter power
spectrum and the CMB temperature and polarization anisotropy power
spectra based on the same parameters used in the background world
model. The CMB temperature and polarization anisotropies are
naturally gauge-invariant, and for the matter power spectrum we
present the power spectrum of density perturbation based on the CCG
which is also a gauge-invariant concept; i.e., density perturbation
in the CCG is the same as a unique gauge-invariant combination
between the density perturbation and the velocity perturbation of
the CDM component. Despite the variety of outcome in the
redshift-distance relation in the parameters used (see right-bottom
panel in Fig.\ \ref{fig:BG-evolution}) the matter power spectra of
the $\phi$CDM models are all similar with some tilt relative the
fiducial $\Lambda$CDM model, whereas the differences in the CMB
power spectra are less distinguished. Figure
\ref{fig:pert-evolution-on} shows that when we properly include the
DEP the three gauges give identical results both for the
$\Lambda$CDM and $\phi$CDM cases.

%
%
\begin{figure}
\begin{center}
\includegraphics[width=8.6cm]{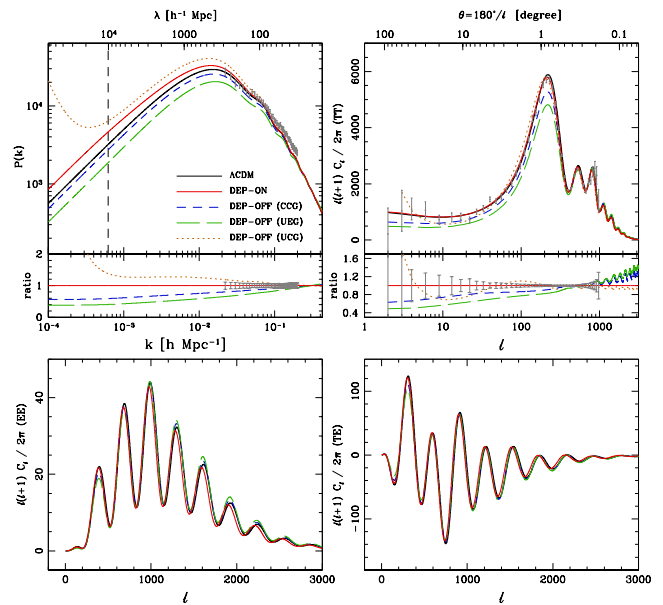}
\caption{
         The same as Fig.\ \ref{fig:pert-evolution-on} now {\it ignoring}
         the DEP (DEP-OFF) in the CCG (blue, dashed), the UEG
         (green, long dashed), and the UCG (brown, dotted curves)
         for $\lambda_2 = 1.0$.
         Red solid curves represent power spectra with proper DEP (DEP-ON).
         The power spectra ignoring the DEP
         apparently depend on the gauge choice which reflects
         internal inconsistency of the system.
         For matter and CMB TT power spectra,
         recent measurements from SDSS DR7 LRG \cite{SDSS-DR7-LRG} and
         WMAP 5-year \cite{WMAP-5yr} data (including the cosmic variance)
         have been added (grey dots
         with error bars), and power ratios between cases ignoring and
         considering DEP are also shown for CCG, UEG, and UCG,
         together with fractional errors of observed spectra.
         }
\label{fig:pert-evolution-off}
\end{center}
\end{figure}

Now, in Fig.\ \ref{fig:pert-evolution-off} we ignored (set equal to
zero by hand) the perturbed part of dark energy. Apparently, the
results depend on the gauge conditions used. As the values of
gauge-invariant variables depend on the gauge conditions used in the
calculation this {\it alarms} inconsistency of the system. Such
differences are expected because by ignoring the DEP the perturbed
system of equations becomes {\it inconsistent}. The presence of
fluctuations in the matter and metric simultaneously and inevitably
excites fluctuations in the dark energy. And it is {\it not} allowed
to turn off the DEP by hand. The issue we would like to address,
however, is whether we could ignore the DEP in practice. Our result
in Fig.\ \ref{fig:pert-evolution-off} shows that ignoring the DEP
easily leads to observationally significant deviations in the power
spectra which are even gauge dependent.

In our normalization the matter power spectrum shows about
$-20$\%/$-34$\%/$+20$\% ($-10$\%/$-19$\%/$+8.9$\%) error caused by
ignoring the DEP at $k \simeq 0.022h\textrm{Mpc}^{-1}$ in the
CCG/UEG/UCG; the values inside parenthesis are for $\Omega_{\phi i}
= 0.0225$ which is one half of the value used in our Figures. The
current observation from SDSS DR7 LRG (Luminous Red Galaxies) shows
$11\%$ (correlated) error at the same scale, which is already
smaller than the deviations caused by ignoring the DEP in all
gauges. The CMB temperature power spectrum shows about
$-9.8$\%/$-18$\%/+.64\% ($-6.0$\%/$-10$\%/+.63\%) error caused by
ignoring the DEP at $\ell=200$; the WMAP 5-year data in this scale
has about $2\%$ (binned) error (mostly due to the cosmic variance);
as the figure shows the small error in the UCG is only due to a
coincidence in this scale.

%
%
\begin{figure}
\begin{center}
\includegraphics[width=8.6cm]{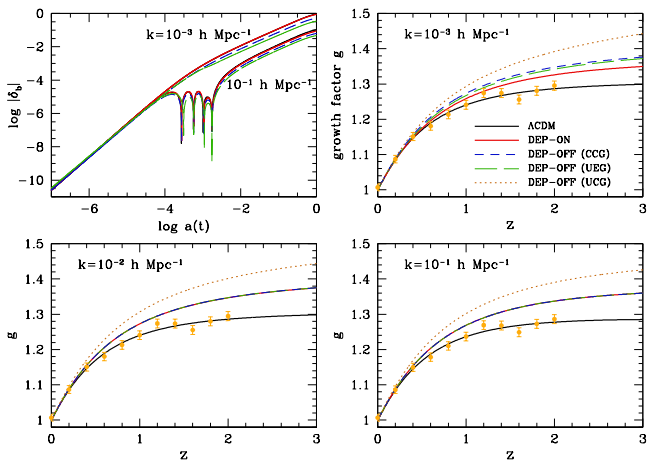} \caption{
         Evolution of baryon density perturbation (top-left),
         and the normalized perturbation growth factor $g \equiv (\delta_b/a)$
         in three different scales for $\lambda_2 = 1.0$;
         due to strong deviations we omit UCG cases in top-left panel.
         As we normalize $g$ to unity at present,
         the effect of DEP appears only in the large scale
         (top-right), and except for the UCG, it has no effects
         in the two small scales (bottom panels).
         We add $1\%$ error bar expected from future X-ray and
         weak lensing observations \cite{X-ray}.
         }
\label{fig:growth-factor}
\end{center}
\end{figure}

Thus, deviations depend directly on the amount of $\Omega_{\phi i}$
in the early scaling era. In our scaling dark energy model by
reducing $\Omega_{\phi i}$ the deviations caused by ignoring the DEP
become proportionally smaller. However, this does not imply that our
model is an extreme example in the effects of DEP. In fact, we can
easily introduce models theoretically (i.e., not by hand) where
$\Omega_{\phi}$ is negligible during the nucleosynthesis era but
becomes significant during later radiation and early matter eras and
then reduces to the dark energy in recent era so that the rest of
the cosmological effects are indistinguishable but the resulting
power spectra in the matter and CMB are substantially different in
diverse ways, see \cite{Park-2009}.

%
%
This still implies that the substantial deviations in the power
spectra due to DEP are mainly caused during the scaling era. This is
partly supported by studying the baryon density perturbation growth
factor $g$ in the recent past which provides another domain where
theory meets with observation, see Fig.\ \ref{fig:growth-factor}. In
the CCG and the UEG cases the DEPs do not cause difference in
observationally relevant small scales. Although the observationally
distinguishable substantial deviations in the UCG case can be
regarded as exceptional peculiarity of that gauge choice, this still
indicates the inconsistency of equations without DEP and potential
danger of ignoring the DEP without proper confirmation. That is, by
ignoring the DEP the system of equations becomes inconsistent and
even (gauge-invariant) observable results depend on the gauge
choice; thus, Fig.\ \ref{fig:growth-factor} shows the particular
importance of taking proper gauge in the absence of the DEP. Notice
that in our realistic situation with early radiation era the growth
factor shows scale dependence.

%
%
In this {\it Letter} we investigated the roles of DEP in a dynamical
dark energy model based on the scalar field. The moral is that when
we consider dynamical dark energy it is essentially important to
take into account the fluctuating aspects of dark energy properly.
When one ignores DEP it is important to show that one can do that
without causing observationally significant differences. Our model
shows an example where it is crucially important to include the DEP.
Otherwise, the system of equations becomes inconsistent, and the
consequent results are not reliable compared with currently
available observations.

%
%
{\bf Acknowledgments:} H.N.\ was supported by grants No.\ C00022
from the Korea Research Foundation (KRF) and No.\ 2009-0078118 from
the Korea Science and Engineering Foundation (KOSEF) funded by the
Korean Government (MEST). J.H.\ was supported by KRF Grant funded by
the Korean Government (MOEHRD, Basic Research Promotion Fund) (No.\
KRF-2007-313-C00322) (KRF-2008-341-C00022), and by Grant No.\
R17-2008-001-01001-0 from KOSEF.

%
%



\begin{thebibliography}{99}
\bibitem{SN}
         A.G. Riess {\it et al}., Astron. J. {\bf 116}, 1009 (1998);
         S. Perlmutter {\it et al}., Astrophys. J. {\bf 517}, 565 (1999).
\bibitem{BAO}
         D.J. Eisenstein {\it et al.}, Astrophys. J. {\bf 633}, 560 (2005).
\bibitem{SDSS-DR7-LRG}
         B.A. Reid {\it et al.}, arXiv:0907.1659v2 (2009).
\bibitem{WMAP-5yr}
         G. Hinshaw {\it et al.}, Astrophys. J. Suppl. {\bf 180}, 225
                    (2009);
         M.R. Nolta {\it et al.}, {\bf 180}, 296 (2009).
\bibitem{X-ray}
         A. Vikhlinin {\it et al}., astro-ph/0903.5320 (2009).
\bibitem{no-DE-pert}
         D. Huterer and E.V. Linder, Phys. Rev. D {\bf 75}, 023519
                    (2007);
         M.J. Mortonson, W. Hu, and D. Huterer, Phys. Rev. D
                    {\bf 79}, 023004 (2009).
\bibitem{Lifshitz-1946}
         E.M. Lifshitz, J. Phys. (USSR) \textbf{10}, 116 (1946);
         W.B. Bonnor, Mon. Not. R. Astron. Soc.
                       \textbf{117}, 104 (1957);
         H. Nariai, Prog. Theor. Phys. \textbf{41}, 686 (1969);
         J.M. Bardeen, Phys. Rev. D \textbf{22}, 1882 (1980).
\bibitem{DE-pert}
         F. Perrotta and C. Baccigalupi, Phys. Rev. D {\bf 59},
                     123508 (1999);
         C. Ma, R.R. Caldwell, P. Bode, and L. Wang, Astrophys. J.
                     {\bf 521}, L1 (1999)
\bibitem{Bassett-etal-2008}
         B.A. Bassett {\it et al}., JCAP {\bf 0807}, 007 (2008)
\bibitem{Kowalski-etal-2009}
         M. Kowalski {\it et al}., Astrophys. J. {\bf 686}, 749 (2008).
\bibitem{HN-scaling-2001}
         J. Hwang and H. Noh, Phys. Rev. D \textbf{64}, 103509 (2001).
\bibitem{Bean-etal-2001}
         R. Bean, S.H. Hansen, and A. Melchiorri, Phys. Rev. D
                  {\bf 64}, 103508 (2001).
\bibitem{HN-CMB-2002}
         J. Hwang and H. Noh, Phys. Rev. D \textbf{65}, 023512 (2002).
\bibitem{Bardeen-1988}
         J.M. Bardeen, {\it Particle Physics and Cosmology}, edited by
                       L. Fang, and A. Zee, (Gordon and Breach, London, 1988),
                       p1;
         J. Hwang, Astrophys. J. \textbf{375}, 443 (1991).
\bibitem{Seljak-1996}
         U. Seljak and M. Zaldarriaga, Astrophys. J. {\bf 469}, 437 (1996).
\bibitem{Park-2009}
         C.-G. Park, J. Hwang and H. Noh, In preparation.
\end{thebibliography}
\end{document}